\documentclass[fleqn,a4paper,12pt]{article}
\usepackage{amssymb}
\usepackage{makeidx}
\usepackage{amsmath}
\usepackage{graphicx}
\usepackage{lscape}
\usepackage{a4}
\usepackage{epsfig}
\usepackage{graphicx}

\renewcommand{\bar}{\overline}
\newcommand{\newc}{\newcommand}
\newc{\beq}{\begin{equation}} \newc{\eeq}{\end{equation}}
\newc{\bea}{\begin{array}} \newc{\eea}{\end{array}}
\newc{\ri}{{\mathrm i}}
\newc{\bW}{{\mathbf W}}
\newc{\bR}{{\mathbf R}}
\newc{\bN}{{\mathbf N}}
\newc{\Psibar}{\overline\Psi}
\newc{\w}{{\bf w}}
\newc{\E}{{\mathbf{E}}}
\newc{\bp}{{\bf p}}
\newc{\ta}{\tilde a}
\newc{\bV}{{\bf V}}
\newc{\bfV}{{\bf V}}
\newc{\bfG}{{\bf G}}
\newc{\bx}{{\bf x}}
\newc{\bu}{{\bf u}}
\newc{\bP}{{\bf P}}
\newc{\bJ}{{\bf J}}
\newc{\bK}{{\bf K}}
\newc{\pd}{{\partial}}
\newc{\ti}{{\times}}
\newc{\bA}{{\bf A}}
\newc{\bE}{{\bf E}}
\newc{\bfn}{{\bf\nabla}}
\newc{\ho}{\hookrightarrow}
\newc{\ra}{\rightarrow}
\newc{\bv}{{\bf v}}
\newc{\bb}{{\bf b}}
\newc{\bc}{{\bf c}}
\newc{\bd}{{\bf d}}
\newc{\tbb}{\tilde{\bf b}}
\newc{\tbc}{\tilde{\bf c}}
\newc{\tbd}{\tilde{\bf d}}
\newc{\bz}{{\bf 0}}
\newc{\bun}{{\bf 1}}
\newc{\bL}{{\bf L}}
\newc{\bS}{{\bf S}}
\newc{\bB}{{\bf B}}
\newc{\br}{{\bf r}}
\newc{\sig}{{\mathbf\sigma}}
\newc{\eg}{{\it e.g.\ }}
\newc{\bpi}{{\mathbf\pi}}
\newc{\ie}{{\it i.e.\ }}
\newc{\etal}{{\it et al}}

\catcode`@=11 \long
\def\@caption#1[#2]#3{\par\addcontentsline{\csname
ext@#1\endcsname}{#1} {\protect\numberline{\csname
the#1\endcsname}{\ignorespaces #2}} \begingroup \small
\@parboxrestore \@makecaption{\csname fnum@#1\endcsname}
{\ignorespaces #3}\par \endgroup} \catcode`@=12

\begin{document} \begin{titlepage} \vskip 2cm
\begin{center} {\Large\bf  Symmetries and modelling functions
  for diffusion processes}

\vskip 3cm {\bf  A.G. Nikitin$^a$,
S.V. Spichak$^a$, Yu. S.  Vedula$^b$  and A.G. Naumovets$^b$}
\vskip 5pt
 {\sl $^a$Institute of
Mathematics of National Academy of Sciences of Ukraine,\\ 3
Tereshchenkivs'ka Street, Kyiv-4, Ukraine, 01601
\\e-mail: nikitin@imath.kiev.ua;
  \vskip 2pt
 {\sl $^b$Institute of Physics of National Academy of
 Sciences of Ukraine, \\ 46 Prospect Nauki, Kyiv-28,
 Ukraine, 03028\\e-mail: naumov@iop.kiev.ua}}
\end{center}
 \vskip .5cm \rm
{\bf Short title}: Symmetries and modelling functions
\vskip .5cm \rm

\vskip .5cm \rm
\begin{abstract}A constructive approach to theory of diffusion
processes is proposed, which is based on application of both the
symmetry analysis and method of modelling functions. An algorithm
for construction of the modelling functions is suggested. This algorithm is based on the error functions expansion (ERFEX) of experimental concentration profiles. The high-accuracy analytical description of the profiles provided by ERFEX approximation allows a convenient extraction of the concentration dependence of diffusivity from experimental data and prediction of the diffusion process. Our analysis is exemplified by its employment to experimental results obtained for surface diffusion of lithium on the molybdenum (112) surface pre-covered with dysprosium. The ERFEX approximation can be directly extended to many other diffusion systems.
 \end{abstract}

\end{titlepage}

\section{Introduction}

Experimental and theoretical studies of diffusion processes are of a
great importance for various branches of physics, biology, chemistry
and other natural sciences. In addition, such studies have important
applications in medicine and many technological processes. A special
interest is exited by surface diffusion processes which appear in many physical and chemical systems. In
particular, they are used in various kinds of nanotechnologies.

The theory of diffusion processes started in 1855 when Fick
derived his classical diffusion equation \cite{fick}
\beq\label{fick} \frac{\partial \theta}{\partial t}-\frac{\partial
}{\partial x_a} \left(D\frac{\partial \theta}{\partial x_a}\right)=0,\eeq
which still is a corner stone of the diffusion theory. In equation
(\ref{fick}) $D$ is a diffusion coefficient, in general case depending on species
concentration $\theta$, and $x_a$ with $a=1,2,3$ are spatial variables (summation over the repeated indices $a$ is imposed). Being supplemented by an appropriate initial data, equation (\ref{fick}) serves as a background for description of such
diffusion processes which are characterized by diffusion flows linear in concentration gradients and not depending explicitly on space
and time variables.

Two standard problems of a diffusion theory are:

1) To describe time evolution of the diffusion process, and

2) To specify the dependence of the diffusion coefficient on concentrations of diffusing species.

Of course, these problems are closely related, since if we know how
the diffusion coefficient depends on concentration $\theta$, then the time
evolution of the corresponding diffusion process can be found using the Fick equation (\ref{fick}) and the related initial data. On the
other hand, if we know $\theta$ as a function of time variable $t$ and spatial variables $x_a$, then we can find $D$ solving the inverse
diffusion problem using again equation (\ref{fick}).

Both mentioned problems are very complicated and in general need rather sophisticated techniques. Even if we know the diffusion
coefficient as an explicit function of concentration, then generally speaking it is possible to find only an approximate (numerical)
solution of the first problem if at all. The second problem has a much more complex character, but in the case of a sharp step-like initial $\theta$ profile it is possible to use the
Boltzmann-Matano (BM) approach \cite{matano}
and reconstruct the concentration dependence $D(\theta)$ of the diffusion coefficient. This approach
 enables one to make a numerical calculation of the diffusion coefficient,
 but its accuracy is not very high, especially for small and large concentrations $\theta$.

Experimental data and numerical solutions are very important for description
of a diffusion process, but to formulate its theory it is desirable to
create some analytical expressions for studied values. Unfortunately,
there are only few known exactly solvable realistic diffusion problems,
the most famous of them is probably the Barenblat one \cite{Barenblat}. Thus it is
a common practice to use rather rough analytic presentations of $D(\theta)$ to
make a qualitative analysis of diffusion process (see, e.g., \cite{routh}).

In the present paper we propose a new method for description of time
evolution of a diffusion process and calculation of the diffusion
coefficient. The distinct  feature of our approach is that we find
both functions $\theta=\theta(t,x)$ and $D=D(\theta)$ in an explicit
form, i.e., solve both problems 1) and 2) analytically. To achieve
this goal we start with experimental data for a particular diffusion
system and make  the error functions expansion (ERFEX) of
concentration profiles. Analytic description of diffusion processes
is very convenient for their qualitative analysis. Moreover, our
description appears to be rather good quantitatively also; its
deviation from experimental data does not exceed the inaccuracy of
measurements.

 We apply this approach to describe in detail the surface diffusion of Li deposited on the molybdenum (112) surface which had been previously covered with a submonolayer of dysprosium. One more process, the diffusion of Dy adsorbed on Mo(112), is used to examine the method generality. Moreover, we  believe that it can have a  much wider application area.

\section{Experimental data and symmetries}

Let us start with experimental data representing surface diffusion of Li on the Mo(112) surface precovered with a 0.25 monolayer of dysprosium (below we designate it as Dy-Mo(112) surface). The data were obtained in ultra-high vacuum using local measurements of the work function by a contact potential method (see \cite{vedula}-\cite{ved2} for details).

\begin{figure}[th]
\centerline{\includegraphics[width=100mm]{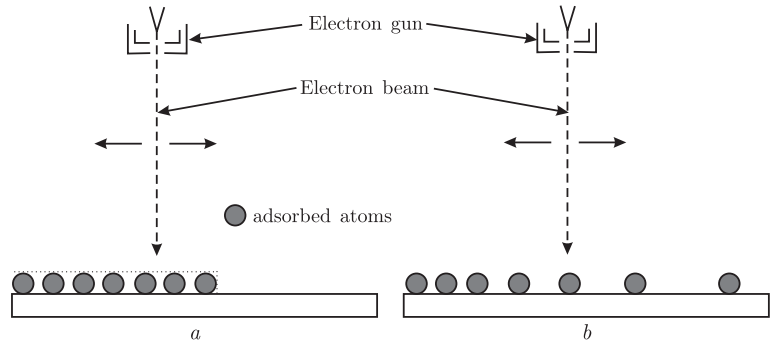}}
\caption {Probing the surface distribution of adatoms by scanning contact potential microscopy. (a) The initial step-like adatom distribution. (b) The adatom distribution after surface diffusion.}\label{f0}
\end{figure}

A schematic sketch of the method, termed scanning contact potential microscopy, is shown in Fig.~\ref{f0}.
At the beginning of the experiment, we uniformly covered the clean surface of a (112) oriented Mo single crystal with dysprosium (its surface density amounted to 0.25 of a monoatomic layer) and equilibrated it by annealing at T=1100 K. The Dy adatoms served thereafter as a controllable admixture which could affect the diffusion kinetics of lithium [6]. Then, using a semiplane mask (screen) placed between the Li evaporator and the prepared substrate, a half of the crystal surface was covered at room temperature with Li while its another half remained clean of Li
(Fig. 1,a).  The surface
 was scanned with an electron beam formed by an electron gun.
 The movable beam was used to record the distribution of the
 contact potential (work function) over the sample surface
 by P.A.~Anderson's (retarding field) method.
 In separate experiments, the work function change was carefully
 calibrated with respect to the absolute surface concentration of adsorbed atoms (adatoms). The calibration data served to convert
 the work function values to adatom concentrations, $n$. To characterize
the relative concentration of adatoms with respect to substrate
surface atoms, we shall use, as it is conventional in surface
science, the term "degree of coverage" (or, for short, "coverage").
It is defined as $\theta=n/n_M$, where $n_M$ is the concentration of
the substrate surface atoms ($n_M=8.3\cdot 10^{14} cm^{-2}$ for the
Mo(112) surface). The coverage $\theta=1$ is usually termed the
 geometrical monolayer.

Under the experimental conditions provided in
\cite{vedula,ved1,ved2}, there was neither evaporation of the
adsorbate into vacuum nor its diffusion (drain) into volume of the
Mo substrate. Thus the experimental results relate to a case a
"pure" surface diffusion which could be described by equation
(\ref{fick}). Notice that the length of the crystal sample along the
diffusion direction was about 10 mm while the extension of the
diffusion zone in the experiments did not exceed  1.2-1.5 mm. Thus
the boundary effects connected with the finite size of the sample
could be neglected

The initial Li distribution was step-like shaped. Then upon
heating  the adsorbate profile spreads out due to surface diffusion.
Since the edge of the step was oriented normally to the atomic
channels on the Mo(112) surface, the diffusion proceeded
quasi-one-dimensionally along the channels, i.e. along the
[${\bar 1},{\bar 1},1$]  direction [5-7].

The experiment consisted of a series of measurements in which we
recorded the time evolution of the coverage profiles due to
diffusion at a constant annealing temperature. At the beginning of
each experiment (t=0), a standard (step-like) initial coverage
profile of the adsorbate was created and recorded on the crystal
kept at room temperature, at which the adatom mobility is
negligible. This profile is labeled t=0. Then the crystal sample was
annealed at a fixed temperature. From time to time, the annealing
was interrupted and the crystal quickly cooled down to room
temperature to record a new coverage profile arising due to
diffusion. After that the annealing was continued, and so on. In
this way we obtained a series of profiles corresponding to different
annealing times and a constant annealing temperature. Such
experiments could be repeated at different annealing temperatures to
determine the temperature dependence of the diffusion kinetics.

\begin{figure}[th]
\centerline{\includegraphics[width=80mm]{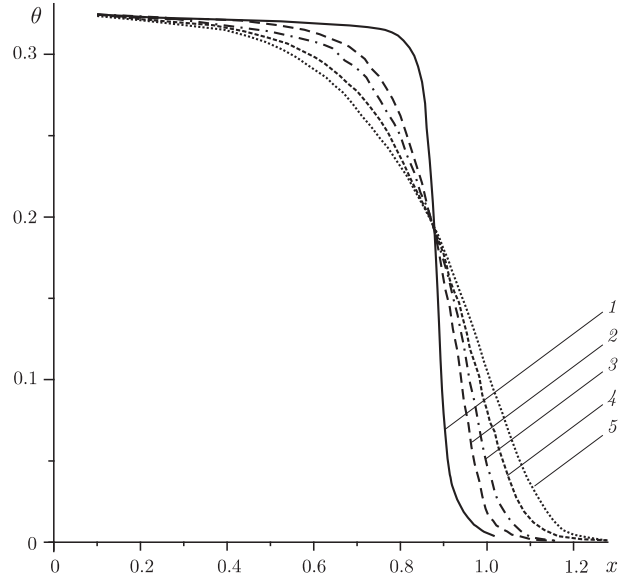}}
\caption{Coverage profiles of Li adsorbed by Dy-Mo(112) at T=600 K: initial,
$t=0$ (1), and measured at $t=1200 $ s (2), $t=2100$ s (3), $t=3600$ s (4), $t=5400$ s (5). The $x$-
coordinate gives distance in mm.} \label{f1}
\end{figure}

Measurements were made at times $t=0, 1200, 2100, 3600$ and $5400$
(seconds) at stable temperature $T=600 K$,
experimental error in $\theta$ was $\Delta \theta=0.003$.
The recorded coverage profiles are presented in Fig.~\ref{f1}.

The initial profile ($t=0$) is step-like shaped, but technologically
it was impossible to form an ideal step. The profiles obtained as a
result of surface diffusion have a common intersection point at
$x=0.88, \theta=0.192$ and have rather smooth contours. Moreover, a
convention is used to set $\theta=0$ in the region where its value
is below the measurement accuracy.

Our task is to give a phenomenological theory of the related
diffusion process. Abstracting from complicated underlying physical effects
which are discussed in paper
\cite{naum}, we will describe the time evolution of the
diffusion process and derive the dependence of the diffusion coefficient on concentration.

To achieve our goal, we will exploit symmetries which form the
integral part of diffusion processes. A precise analysis of the
experimental data makes it possible to find a specific symmetry
which characterizes coverage profiles. Namely, let us fix a profile
$\theta$ recorded at time $t$ and consider it as a given function of
$x$. Then any other profile $\theta'$ measured at time $t'$ can be
obtained from $\theta$ using the following transformation:
\beq\label{3}x\to x'=(x-x_0)\sqrt\frac{t}{t'}+x_0,  \eeq where
$x_0=0.88$ is coordinate of the point common for all coverage
profiles.

This statement can be verified directly or using  computer fits to
the experimental curves. For example, starting with experimental
data for $t=3600$ s and applying transformation (\ref{3}), one can
reproduce profiles for $t=1200 \texttt{ s}, \ 2100 \texttt{ s}$ and
$5400 \texttt{ s}$.

 It should be stressed that the found symmetry is rather exact.
 As a rule, a deviation of curves obtained
 using transformation (\ref{3})
 from experimentally measured  profiles
is within the limits of experimental error, and this deviation decreases with growing time.
For example, comparing experimental data for
 coverage profile at $t=5400$ s (Fig.~\ref{f1}) and the
 corresponding values obtained via change (\ref{3}) we conclude
 that they are very similar, namely, the differences are below the experimental error.

\section{Time evolution derived from symmetries}

The exact meaning of the symmetries discussed in Section 2 is that
functions $\theta(t,x)$ describing coverage profiles are invariant with
respect to the following one-parametric group of transformations:
\beq\label{trans0}t\to t'=e^{2\alpha}t, \ x\to
x'=xe^\alpha+x_0(1-e^\alpha),\eeq where $\alpha$ is a real parameter.
Indeed, solving the first of equations (\ref{trans0}) for $e^\alpha$
and using the second equation we come to relations (\ref{3}).

Starting with (\ref{trans0}) and using tools of the classical group
analysis, it is possible to describe time evolution of profile
$\theta(t,x)$. Indeed, the infinitesimal operator of transformations
(\ref{trans0}) has the form:
\[
X=\eta\frac{\partial}{\partial
t}+\xi\frac{\partial}{\partial x}\equiv 2t\frac{\partial}{\partial
t}+x\frac{\partial} {\partial x}-x_0\frac{\partial}{\partial x},
\]
where  $\eta=\frac{\partial t'}{\partial
\alpha}|_{\alpha=0}$ and $\xi=\frac{\partial x'}{\partial
\alpha}|_{\alpha=0}$ \cite{olver}.
 The invariance of coverage profiles with respect to transformations
 (\ref{trans0}) means that $\theta(t,x)$ solves the following equation
 \cite{olver}:
 \beq\label{inv_cond}X\theta(t,x)=0.\eeq

It follows from (\ref{inv_cond}) that time evolution of coverage
profiles $\theta(t,x)$ is described by the following equation:
\beq\label{te}\frac{\partial \theta}{\partial t}=\frac{x-x_0}{2t}
\frac{\partial \theta}{\partial x}\ \ \ \texttt{or }\ \
\frac{\partial \tilde\theta}{\partial t}=\frac{y}{2t} \frac{\partial
\tilde\theta}{\partial y},\eeq where $y=x-x_0$,
$\tilde\theta(t,y)=\theta(t,y+x_0)$. As an initial condition we can
choose one of measured profiles, say that one which corresponds to
$t=t_2=1200$:
\beq\label{ic}\tilde\theta(t_2,y)=\theta_2=\theta_2(y+x_0)\eeq where
$\theta_2(x)$ is a function given numerically in Table 1. Henceforth
we omit tilde and write $\theta(t,y)$ instead of $\tilde
\theta(t,y)$.

\begin{figure}[th]
\centerline{\includegraphics[width=80mm]{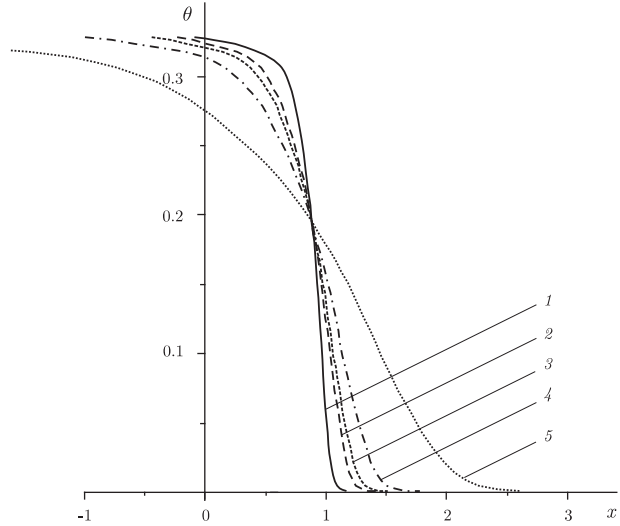}}
\caption{Coverage profiles $\theta(t,x)$ for Li on Dy-Mo(112) at T=600 K: experimental for $t=2100$
s (1)
 and theoretical for $t=7000$ s (2), 10000 s (3), 20000 s
(4), 100000 s (5). The $x$-coordinate gives distance in mm.}
\label{f3}
\end{figure}

 Thus we can describe the time evolution of the coverage profiles without a diffusion equation. Solving problem
(\ref{te}) with condition (\ref{ic}) and using numerical data given in the Appendix
we can find shapes of these profiles for any time.
Some of such profiles are presented in Fig.~\ref{f3}.

As we see, the very existence of the symmetries in experimental data
makes it possible to predict shapes of the coverage profiles which
will appear at various times of heating. This statement is valid for
any diffusion system which admits symmetries (\ref{trans0}). But if
we are interested in analytical description of the diffusion
process, we have to pose initial conditions analytically. A basic problem is to
create a consistent model of the diffusion process, i.e. find the
dependence of the diffusion coefficient on concentration, which is
important for physical interpretation.

In the following sections we solve this problem and give explicit
representations of coverage profiles $\theta(t,x)$ in analytical form.

\section{Symmetries of diffusion models}

 Transformations (\ref{3}) have a stable point $x=x_0=0.88$ and
 $\theta=0.194$ which was the only one not being changed.
 It is convenient to choose just $x_0$ as a zero point of our coordinate system,
 i.e. to use variable $y=x-x_0=x-0.88$
 instead of $x$.

  Taking into account
that in fact we deal with a process which is one-dimensional with
respect to spatial variables, it is possible to reduce equation
(\ref{fick}) to the following form: \beq\label{fick1}\frac{\partial
\theta}{\partial t}-\frac{\partial }{\partial y}\left(
D\frac{\partial \theta}{\partial y}\right)=0\eeq where $ D=D(
\theta),\ y=x-x_0$.

Formula (\ref{fick1}) represents a rather complicated non-linear
partial differential equation. In addition, we do not know the
$\theta$-dependence of the diffusion coefficient $D$. Fortunately,
this equation has a very useful symmetry with respect to scaling of
independent variables, being invariant with respect to
transformations (\ref{trans0}), or \beq\label{trans1}t\to
t'=e^{2\alpha}t, \ y\to y'=ye^\alpha,\ \theta\to\theta'=\theta.\eeq

Just this nice property of the diffusion equation makes it possible
to use the Boltzmann variable $\displaystyle\xi=\frac{y}{\sqrt{t}}$
and search for its similarity solutions $\theta=\theta(\xi)$ where
both $\theta$ and $\xi$ are invariants of transformations
(\ref{trans1}).

Notice that equation (\ref{fick1}) is invariant also with respect to
shifts of independent variables \beq\label{trans5}t\to t+b$, \ \
$y\to y+k\eeq with arbitrary real parameters $b$ and $k$. This
invariance allows one to choose arbitrary initial time and justifies
the transition from $x$ to $y$ which we made above. For some
particular functions $\tilde D(\theta)$, symmetry of the Fick
equation is more extended. A complete classification of symmetry
groups of equation (\ref{fick1}) has been made by Ovsiannikov
\cite{Oves}, a complete group classification of systems  of two
diffusion equations with source terms can be found in
\cite{kniazeva} and \cite{Nik}.

However, symmetries (\ref{trans1}) should be compatible also with the
initial data of our problem, which are of the form:
 \beq\label{t=0}\bea{l}\theta(0,y)=\theta_1(y)\eea,\eeq
 where $\theta_1$ is the initial coverage profile ($t=t_1=0$) represented
 numerically in the Appendix and graphically in Fig.~\ref{f1}.

 We see that the experimentally created profile at $t=0$
 is not strictly step-shaped, and so
 {\it is not invariant with respect to scaling (\ref{trans1})}. However,
 the profile $\theta_1$ is not far from a step and can be
 considered as a perturbed Heaviside function $H(-y)$ multiplied by $0.327$ ($\theta_{max}=0.327$ is the maximum coverage in the initial
 $\theta$ profile).
 On the other hand, and it is an experimental fact, for sufficiently
 large times $t$ solutions of our problem indeed are invariant with respect
 to transformations  (\ref{3}). And if we apply
 this transformation to infinitely small $t'$, all profiles $\theta_1, \theta_2...\theta_4$
 tend to a step-shaped one.
 So we have a direct experimental confirmation of the known mathematical
fact that similarity solutions of the diffusion equation
(\ref{fick1}) can serve as attractors for other solutions. In
other words, the yet unknown diffusion coefficient should depend
on $\theta$ in such a way that the related Cauchy
  problem (\ref{fick1}), (\ref{t=0})
be asymptotically stable with respect to small perturbations of the
initial data (see \cite{Achieser} for exact definitions).

Thus, instead of the actual initial data presented in the
Appendix and Fig.~\ref{f1}, we can consider an idealized situation
when the initial coverage has a step shape, and to suppose that for
$t=0$ the concentration is proportional to the Heaviside function
$H(-y)$: \beq\label{8} \theta(0,y)= \theta_\texttt{max}\cdot
H(-y)=\left\{{\bea{l} 0.327,\ y<0,\\0 ,\ y\geq0 \eea}\right.\eeq
where $\theta_\texttt{max}=0.327$ is the maximum concentration in the
initial coverage profile.

The initial-value problem (\ref{fick1}), (\ref{8}) is invariant with
respect to transformations (\ref{trans1}), and so it is possible to
search for invariant solutions $\theta=\theta(\xi)$ depending on
invariant variable $\displaystyle\xi=\frac{y}{\sqrt{t}}$. In this
way the problem is reduced to the following one:
\beq\displaystyle\label{9}\bea{l}\xi \displaystyle\frac{\partial
\displaystyle\theta }{\partial\xi}+ 2\frac{\partial}{\partial
\xi}\left(
\displaystyle D\frac{\partial \theta} {\partial\xi}\right)=0,\\
\displaystyle\theta(-\infty)=0.327,\ \ \ \theta(+\infty)=0.\eea\eeq

Remember that we do not know yet the dependence of the diffusion
coefficient $D$ on degree of coverage $\theta$. Using experimental
data describing dependence of $\theta$ on $x$ at fixed heating time
$t$ and applying the BM approach \cite{matano} it is
possible to calculate $D$ numerically. Unfortunately, such
calculations cannot be done with a sufficiently good accuracy,
especially in the region of small concentrations. In addition, in
this way we cannot find the diffusion coefficient in an analytical
form. In Section 6 we suggest another approach which presupposes
direct analytical modelling of profiles $\theta(\xi)$.

\section{Generalized diffusion equation }

Thus we had formulated a possible model of the analyzed diffusion process which is based on equation (\ref{fick1})
whose solutions should satisfy the idealized initial conditions (\ref{8}).
However, the possibility of using
Fick equation (\ref{fick1})  is nothing but a supposition which needs additional justifications. In particular,
it is necessary to be ensured that the diffusion flow is linear in the concentration gradient.

However, we can use the well justified fact, i.e., the invariance of
experimental coverage profiles with respect to transformations
(\ref{3}), and set the following problem: to find the most general
evolution equation which is compatible with this symmetry.

Thus let us suppose that the evolution equation admits symmetries
(\ref{3}) and also shifts (\ref{trans5})
(i.e., does not depend explicitly on space and time variables). Then
using tools of classical Lie analysis \cite{olver}, we easily find
its general form: \beq\label{gen}\frac{\partial P}{\partial
t}-\frac{\partial }{\partial y}\left(\tilde D\frac{\partial
P}{\partial y}\right)+G\left(\frac{\partial P}{\partial
y}\right)^2=0\eeq where $P$ is a dependent variable whose evolution
we need to describe, $\tilde D$ and $G$ are arbitrary functions of
$P$. In particular $P$ can represent a degree of coverage, in this
case it should be changed to $\theta$.

We do not present the related routine calculations since our rather
strong restrictions (equation should be of evolutionary type and
admit the above mentioned symmetries) in fact reduce the problem
of deducing of (\ref{gen}) to direct use of the dimension analysis.

Equation (\ref{gen}) is a direct generalization of the Fick equation
(\ref{fick1}), and the latter corresponds to a particular choice $G=0$.

Thus starting with symmetries (\ref{3}), which can be found in the experimental data, and symmetries (\ref{trans5}),
 which are natural transformations for the considered diffusion system, we deduce the generalized  diffusion
 equation (\ref{gen}) which includes the Fick equation (\ref{fick1}) as a particular case.
Possible physical motivations for generalization of equation (\ref{fick1}) to (\ref{gen}) are discussed in Section 10.

\section{Modelling functions for coverage profiles}

It is well known that in the case when diffusion
coefficient $D$ is independent on concentration, the
general solution of the problem (\ref{fick1}), (\ref{8}) is given by the following equation:
\[\theta(t,x)=a \texttt{erfc}(b\xi)\]
where $a= \theta_\texttt{max}/2, b=1/(2\sqrt{D})$, $\xi=x/\sqrt{t}$ and erfc is the
complementary error function:
\[\texttt{erfc}(z)=1-\texttt{erf}(z),\ \ \
\texttt{erf}(z)=
\frac2{\sqrt{\pi}}\int_0^z\texttt{e}^{-t^2}dt.\]

This fact suggests using just the error function as a constructive
element of concentration profiles for $\theta$-dependent diffusivities.
This idea appears to be very successful for description of diffusion processes in general and the
processes discussed in Sections 2 and 9 in particular.

In this section we present and discuss examples of modelling functions for the coverage profiles.
An algorithm for constructing such functions is given in the following section.

\begin{figure}[th]
\centerline{\includegraphics[width=80mm]{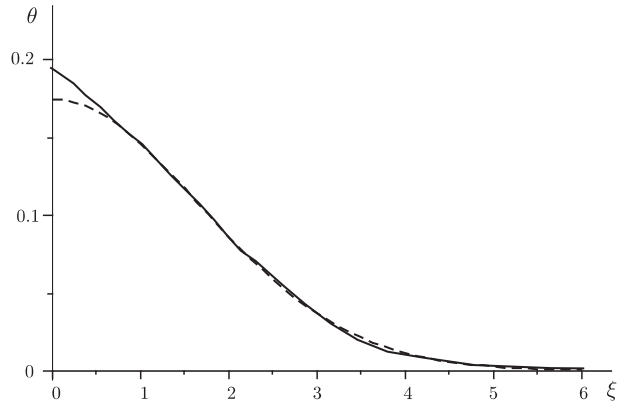}}
\caption{Experimental profile  $\theta(\xi)$ for $t=5400$ (full curve)
and the curve $\theta_1$ given by relation (\ref{sim1})
 (broken curve). Units for $\xi$ are $10^{-3}$ (mm/s$^{1/2})$}
\label{fig02}
\end{figure}

First we consider a rather straightforward representation of
profiles $\theta(\xi)$ which, however, is valid only in a reduced
interval of the Boltzmann variable $\xi$: \beq\label{sim1}\theta=
\theta_{(1)}= a_1(1- \texttt{erf}( b_1\xi)^2)\ \eeq where $a_1$ and
$b_1$ are parameters. Asking for minimal mean-square deviation of
function (\ref{sim1}) and using MAPLE tools we fix parameters $a_1$ and $b_1$ to be
\[ a_1=0.175,\ \
b_1=0.375.\] Function $\theta_{(1)}$ perfectly describes the shape
of profile $\theta(\xi)$ for $\xi$ lying in the interval
$0.6<\xi<\infty$, see Fig.~\ref{fig02}.
In this interval the discrepancy
$|\theta- \theta_{(1)}|$ does not exceed  inaccuracy of measurements. However, for $\xi<0.6$ this
discrepancy increases.

In order to obtain a modelling function for all non-negative $\xi$, it
is sufficient to add a small extra term to $ \theta_{(1)}$ and define:
\beq\label{sim2} \theta_{(2)}= \theta_{(1)}+ a_2\texttt{erfc}(b_2\xi),\ \
 \xi\geq0\eeq
 where $a_2=0.02$ and $b_2=1.7$.

\begin{figure}[th]
\centerline{\includegraphics[width=80mm]{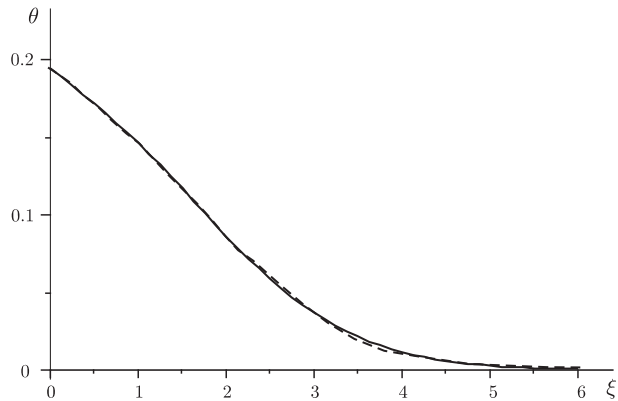}}
\caption{Coverage profiles $\theta(\xi)$: experimental data
  for $t=3600$ s (full curve) and
function  given by relation (\ref{sim2}) (broken curve).
Units for $\xi$ are $10^{-3}$ (mm/s$^{1/2})$}\label{f5}
\end{figure}

Function (\ref{sim2}) gives a very precise presentation for the
profile obtained experimentally at $t=3600$ s; the deviation
$|\theta-\theta_{(1)}|$ does
not exceed the inaccuracy of measurements.
Fig.~\ref{f5}
presents the experimental curve and curve defined by equation
(\ref{sim2}), and it is seen that they practically coincide. More
exactly, the deviation  of values of function
(\ref{sim2}) from experimental data is less than $0.003$.

 One more possible modelling function is
given by the following equation: \beq\label{12} \theta_{(3)}=
a'_1(1-\texttt{erf}(b'_1\xi)^3)+a'_2\texttt{erfc} (b'_2\xi)
\eeq where $a'_1=0.145, b'_1=0.395, a'_2=0.051, b'_2=0.795$.

The modelling function (\ref{12}) is a bit more exact but also more
complicated. Its  deviation from the experimental data for $t=5400$
s is less than $0.0015$, i.e., twice less than the experimental
error.

Thus we wind the modelling functions (\ref{sim2}), (\ref{12})
which
can be obtained starting with {\it a priori} supposition that the
coverage profiles can be described by second or third order
polynomials of error functions and demanding minimal
root-mean-square deviation of these polynomials values  from the
coverage profiles. This supposition is not necessarily valid for other
diffusion systems, e.g., it does not give a constructive way to
build approximate modelling profiles for the system discussed in
Section 9.

In the following section we present a regular way to calculate
modelling functions for any diffusion system using the error
function expansion (ERFEX).

\section{An algorithm for calculation of modelling\\ functions}

We have shown above that erfc functions can be successfully
used as construction elements of the modelling functions of
coverage profiles  at least for a particular  diffusion process.
Now we shall give a regular way for constructing such functions which has a much more extended application area.

Of course a concrete form of the modelling functions strongly
depends on the diffusion system, and we cannot propose a universal
method how to obtain the most simple and exact analytical form of an
arbitrary concentration curve. Nevertheless, in this section we give
an algorithm for calculation of the modelling functions which
can be applied to any sufficiently smooth concentration profile in the diffusion zone.

In general case we propose to use ERFEX and search for modelling
functions in the form \beq\label{mf5}\theta(\xi)=\sum_{i=1}^n
A_i\texttt{erfc} (k_i(\xi-\xi_i))\eeq where $\xi_i\ (i=1,2,...n) $
are some fixed values of Boltzmann coordinate, $k_i$ and  $A_i$ are
parameters which should be specified.

Let $\xi_{i-1}<\xi_i$, or $\xi_1<\xi_2<\cdots<\xi_n$, then optimal values
of $k_i$ lie inside the interval
\beq\label{ki}\frac{0.25}{\xi_{i+1}-\xi_i}\leq k_i\leq
\frac{1}{\xi_{i+1}-\xi_i}. \eeq

In particular it is possible to choose the points $\xi_1,\xi_2\cdots,\xi_n$ in a regular way, i.e., with the same
distances $\xi_{i+1}-\xi_i$ for all $i$, and to fix all parameters to be
$k_1=k_1=\cdots=k_n=p$ with some $p$ compatible with (\ref{ki}).

Let $\theta_1,\theta_2...\theta_n$ be known values of coverage at
points $\xi_1,\xi_2,...,\xi_n$ and $M$ be a matrix whose elements
are $M_{ij}=\texttt{erfc}(k_i(\xi_i-\xi_j))$. Then parameters $A_1,A_2,...,A_n$
are easily found by solving the following system of linear algebraic equations:
\beq\label{eq}M_{ij}A_j=\theta_i,\ i=1,2,\cdots n,\eeq
where summing up over the repeated index $j$ is imposed
from $j=1$ to $j=n$. Equations (\ref{eq}) are nothing but
relation (\ref{mf5}) considered at points $\xi_1,\xi_2,...,\xi_n$.

Let us apply the algorithm to find a modelling function for the
diffusion process discussed in Sections 2-6. Consider the results
presented in the Appendix for $t~=~5400$~s. For simplicity we choose
points $\xi_1, \xi_2,..$ in a regular way. Namely, we chose in the
table selected data corresponding to $x=0.63$, $0.68$, $0.73$,
$0.78$, $0.83$, $0.88$, $0.93$, $0.98$, $1.03$ and $1.118$ (remember
that $x_0=0.88$), and fix $k_i=0.7$ for all $i=1,2,...n \ (n=12)$.
At that, system (\ref{eq}) is
 easily solved with using symbolic calculus
program MAPLE.
 As a
result we find the following representation for $\theta(\xi)$:
\beq\bea{l}\label{BB} \theta(\xi) =
-0.0086\texttt{erfc}(0.7\xi+3.8865)+0.0208\texttt{erfc}(0.7\xi+3.2333)\\
-0.0111\texttt{erfc}(0.7\xi+2.5801)+0.0198\texttt{erfc}(0.7\xi+1.9269)\\
+0.0103\texttt{erfc}(0.7x+1.2737)+0.0126\texttt{erfc}(0.7x+0.6205)\\
+0.0242\texttt{erfc}(0.7\xi-0.0359)+0.0328\texttt{erfc}(0.7\xi
-0.7544)\\+0.0385\texttt{erfc}(0.7\xi-1.4730)+0.0211\texttt{erfc}(0.7x-2.1915)\\
-0.0048\texttt{erfc}(0.7\xi-2.9000)+0.0010\texttt{erfc}(0.7\xi-3.6285).\eea\eeq

This a bit cumbersome formula appears to be very precise;
 in the interval $-6<\xi< 6$ it reproduces experimental data
 (presented in the Appendix) with an accuracy not worse than ,
 moreover, for the majority of experimental points this accuracy is not
 worse than
 $0.001$.
 Moreover, using all points given in the Appendix for a fixed time $t$,
 it is possible to find
"the most exact" modelling functions $\theta_E$ whose values simply
coincide with the experimental curve. However, taking into account
the value of experimental error, such business does not seem to be
reasonable.

Notice that using the algorithm with a non-regular distribution of
points $\xi_i$ it is possible to find a more simple form of the
modelling function:
\beq\label{mf1}\theta(\xi)=\sum_{i=1}^4A_i\texttt{erfc}
(R\xi+B_i),\eeq where \beq \label{mf2}\bea{l}A_1= 0.103,\
A_2=0.234,\ A_3=0.114, \ A_4=0.044, \\ B_1=-1.87,\ B_2=-0.73, \
B_3=1.154,\ B_4=3, \ R=0.64.\eea\eeq

Being much more simple than (\ref{BB}), function (\ref{mf1}) is
rather exact too; the corresponding  standard quadratic deviation
from experimental results is less than $0.003$.

\section{Calculation of diffusion coefficient}

Thus we have at our disposal analytic expressions  for  a close approximation of coverage profiles.
Using them we can calculate the
diffusion coefficient $D(\theta)$ by direct integration of equation
(\ref{9}).

First using the simplest modelling function (\ref{sim1}) we find $D(\theta)$ for concentrations $\theta<0.16$, the
related values of $\xi$ satisfy $\xi>0.6$.
 Substituting (\ref{sim1}) into (\ref{9}) and integrating
 the resultant expression from  $\xi=z>1$ to infinity,
 we obtain
\beq\bea{l}\displaystyle\left.\frac{2 a_1}{\sqrt{\pi}
b_1}\left(\texttt{erf}(b_1\xi) \texttt{e}^{-(
b_1\xi)^2}+\frac1{\sqrt{2}} \texttt{erfc}(\sqrt{2}
b_1\xi)\right)\right| ^z_\infty-\left.\left( \frac{8 a_1
b_1}{\sqrt{\pi}}\texttt{erf} ( b_1\xi) {e}^{-(
b_1\xi)^2}D\right)\right|^z_\infty\\\\ \displaystyle = \frac{2
a_1}{\sqrt{\pi}b_1}\left(\texttt{erf} ( b_1 z)
\texttt{e}^{-(b_1z)^2}+4 b_1^2\texttt{erf} ( b_1 z)
{e}^{-(b_1z)^2}D\right)=0.\label{DD1}\eea\eeq Solving (\ref{DD1})
for $D$ and using (\ref{sim1}) we obtain:
\beq\displaystyle\label{D2} D=\frac1{4
b_1^2}\left(1+\frac{\texttt{e}^{Z^2} \texttt{erfc}(\sqrt{2}Z)}
{\sqrt{2(1-\frac{\theta}{ a_1})}}\right),\ \
Z=\texttt{erfinv}\left(\sqrt{1-\frac{\theta}{ a_1}}\right). \eeq
where $\texttt{erfinv}(\cdot)$ is the inverse error function defined
by means of $\theta=\texttt{erf} (\texttt{erfinv}(\theta)).$

Formula (\ref{D2}) defines the diffusion coefficient $D$ as an
explicit function of concentration and is valid for all $\theta$
lying in interval $0\leq \theta<0.16$.

In an analogous way, i.e., by direct integration of equation
(\ref{9}), it is possible to find the diffusion coefficient $D$
starting with other modelling functions $\theta(\xi)$ found in
Sections 6 and 7. The general expression for $D$ is given by the
following equation:
 \beq\label{D}    D(\xi)=\frac{\int_\xi^\infty
z\frac{d \theta(z)}{d z}dz} {2\frac{d  \theta(\xi)}{d \xi}},\eeq
which, together with a modelling function for $\theta(\xi)$,
determines the diffusion coefficient as a function of $\theta$ given
in a parametric form.

In contrast to the BM approach, to find the
dependence of the diffusion coefficient on concentration we simple
need  to calculate a definite integral of the {\it known} function
 and divide it by the redoubled derivative of (known)
function $\theta$ with respect to $\xi$. All these operations are
easily handled using, e.g., MAPLE tools.

In Fig.~\ref{DG} we present a plot of the diffusion coefficient
(\ref{D}) versus the degree of coverage $\theta$ given by relation
(\ref{BB}).
\begin{figure}[th]
\centerline{\includegraphics[width=80mm]{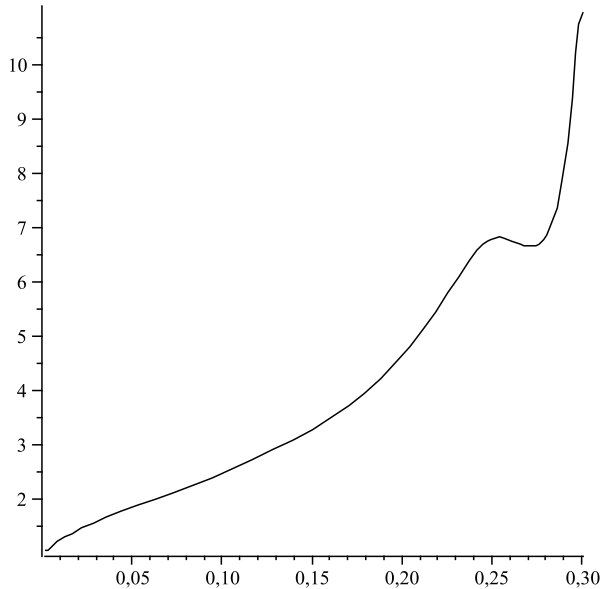}}
\caption{Diffusion coefficient $D(\theta)\ (\texttt{in units }
\texttt{mm}^2/\texttt{s}\cdot10^{-6})$ of Li on Dy-Mo(112)
calculated using modelling function (\ref{BB}). T=600 K.
 } \label{DG}
\end{figure}
These $D$ values are consistent to within 10-25{\%} with the data
obtained in work [6]. However, the maximum
 at $\theta\approx  0.25$ was not revealed in [6], where the graphical
  Matano's evaluation of the coverage profiles was applied.
  This demonstrates that the ERFEX approach provides a more accurate
  processing of experimental diffusion results.

\section{Modelling functions for coverage profiles of Dy adsorbed on
Mo(112)}

We have seen that modelling functions (\ref{BB}), (\ref{mf1})
 give  very
exact analytic expressions for coverage profiles of Li adsorbed on
Mo(112). A natural question arises whether it is possible to find
such functions for modelling the profile shapes for other systems.

\begin{figure}[th]
\centerline{\includegraphics[width=70mm]{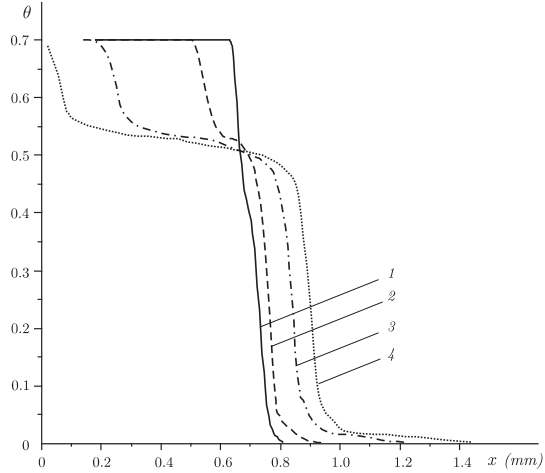}}
\caption{Coverage profile of Dy adsorbed by Mo(112) at T=800 K: initial, $t=0$ (1) and measured at $t=360$ s (2),
$t=2400$ s (3), $t=4800$ s (4) \cite{ved2}. } \label{f21}
\end{figure}

In this section we apply the ERFEX method to another adsorption
system,
namely to Dy adsorbed on Mo(112).
Experimental data for this system were
obtained and discussed in \cite{ved2}. The related plots of
coverage profiles are presented in Fig.~\ref{f21}.

We see that all the profiles recorded in the diffusion
process again have a common intersection point this time at $x=1.69$ which,
however, lies out of the initial profile. Their shapes are much more
specific than ones given in Fig.~\ref{f1}. There are two
zones with a quick change of coverage and three zones where
this change is rather slow. These profiles mirror a structural self-organization in the diffusion region, i.e. formation of a series of two-dimensional adsorbate phases which differ from each other by diffusion parameters and mechanisms \cite{ved2}. Nevertheless, it appears possible
to describe these profiles analytically.

First we represent the coverage profiles using the Boltzmann
variable $\xi=x/\sqrt{t}$ and setting the reference frame $x_0=0$.
As a result we conclude that profiles for $t=2400$ s and $t=4800$ s
became rather close, the mean quadratic deviation of them is less
than 0.003.
Thus it is possible to describe time evolution of profiles using
the approach discussed in Section 3.

\begin{figure}[th]
\centerline{\includegraphics[width=80mm]{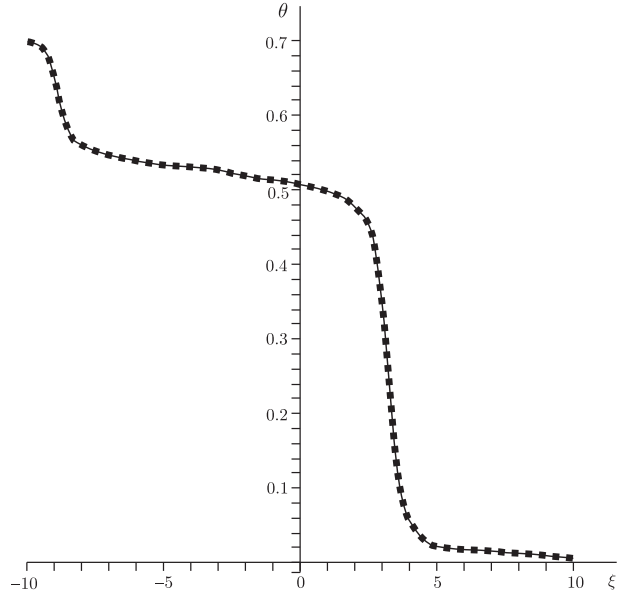}}
\caption{Coverage profile of Dy adsorbed on Mo(112) versus the
Boltzmann variable, measured at $t=4800$ s (full curve) and  profile
described by function (\ref{mf3}) (dotted curve). T=800 K.
 } \label{N}
\end{figure}

Using again the ERFEX, we find the following representation for the coverage profile measured at $t=4800$ s:
\beq\bea{l}\label{mf3} \theta(\xi) = 0.1673\texttt{erfc}(2.2124\xi-6.9912)+
0.06\texttt{erfc}(0.9434\xi-3.3962)\\
+0.0094\texttt{erfc}(0.3334\xi-3.0267)
+0.008\texttt{erfc}(4.4444\xi-8.6222)\\+0.006\texttt{erfc}(2.7778\xi-3.4722)
+0.0058\texttt{erfc}(1.2195\xi)
+0.0095\texttt{erfc}(0.817\xi\\+2.018)+0.0102\texttt{erfc}(0.817\xi+5.1716)
+0.0085\texttt{erfc}(1.9608\xi+15.6862)\\
+0.062\texttt{erfc}(2.5\xi+22.2)
+0.004\texttt{erfc}(1.9608\xi+19.2157).\eea\eeq Function (\ref{mf3})
is a rather precise approximation of the coverage profile
$\theta(\xi)$ at t=4800 s; the mean quadratic deviation from the
experimental results is less than 0.0015. A plot of the related
curves is given by Fig.~\ref{N}.

We see that ERFEX provides a very good analytical representation for a complicated coverage profile of Dy adsorbed on
Mo(112).

\section{Discussion}

The theory of diffusion is both very old and good
developed \cite{Crank}. Nevertheless, it still contains a lot of unsolved
problems which attract attention of numerous investigators.

In the present paper we study three aspects of this theory: using of
symmetries in experimental data to describe the time evolution of a
diffusion process, searching for a generalized Fick equation which
is compatible with these symmetries,
 and construction of modelling functions to describe
concentrations of diffusing substances and calculate
the diffusion coefficient.

Like the diffusion theory, the basic branch of mathematics which
deals with symmetries, i.e.,  the theory of continuous groups, is rather old too. It was started
by Sophus Lie around 20 years after appearance of the Fick theory.
\footnote{The background of the continuous group theory was formulated by S.~Lie in 1875 and published
in 1876 \cite{SophLie}}

The classical group analysis and its modern generalizations present
effective tools for investigation and applications of symmetries of
mathematical models, including diffusion ones. However, to apply these tools
 directly it is necessary to start with a model formulated in terms of
  (partial or ordinary) differential equations.
 \footnote{Integral, difference  and fractional differential
 equations also are subjects of modern group analysis}

A specificity of diffusion systems is that in general we do not
know evolution equation a priory. Even if there is a cogent
argumentation that the process is described by the Fick equation, the dependence of
 diffusion coefficient on concentration is usually unknown. Thus to
apply the Lie analysis it is reasonable {\it to search for symmetries
in experimental data}. And it is the first idea which we use in the
present paper.

The second idea is to use erfc functions as a constructive elements
of  modelling functions for
concentration curves, i.e., to apply the ERFEX expansion. There are two origins of this idea: first,
a similarity to solutions of the Fick equation with a constant
diffusion coefficient, where the erfc function appears very naturally, and secondly, the specific shape of
this function which seems to be ideal for its use as a "brick"
for building typical concentration curves.

Our research is based on a particular diffusion system, i.e., Li
adsorbed on Dy-Mo(112), which is well studied and described in papers
\cite{vedula}-\cite{naum}. This system
represents many basic features of surface diffusion, and thus we
believe that results of our analysis in fact are rather general. A
confirmation of this statement was obtained in Section 9, where we
considered one more surface diffusion system, i.e., Dy adsorbed on
Mo(112).

 A precise analysis
 of the experimental data made it possible to find symmetry
 transformations  (\ref{3}) which connect experimental curves giving
 the coverage profiles. Then we apply these symmetries to describe
 time evolution of the diffusion system and to formulate a possible
 generalization of the Fick equation, given by relation (\ref{gen}), which keeps invariance with
 respect to transformations (\ref{trans0}). A physical motivation
 to search for such generalizations is caused by the fact that the values,
 measured immediately in the
 experiment to judge of the concentration of diffusing species, are not
 necessarily in
 direct proportion to the concentration. Moreover, in some cases we may
 actually be interested not just
 in the concentration of particles, but rather in various physical
 properties connected with it (e.g. electrical
 conductivity , mechanical strength, optical properties, work function etc.).
 Suppose that concentration $\theta$ and the relevant system property $P$
are related by the dependence \beq\label{CC}\theta=F(P).\eeq
Substituting (\ref{CC}) into the Fick equation (\ref{fick1}) we come to
equation (\ref{gen}), where \beq\label{GG} \tilde D(P)=D(F(P)), \ \
G=\tilde D\frac {F''}{F'},\ \ F'=\frac{\partial F}{\partial P}.\eeq
On the other hand, if a diffusion process is accompanied by
dissipation (e.g. due to evaporation or chemical reaction of the
diffusing species), the related mathematical model also needs a
generalization of the Fick equation by inclusion of the terms depending
on $\displaystyle\frac{\partial \theta}{\partial x}$. In order to
create a mathematical model of such process which keeps symmetries
(\ref{trans0}), one should use just the motion equation (\ref{gen})
with $P=\theta$.

We see that the generalized equation (\ref{gen}) appears as a result
of rather straightforward considerations and presents an alternative
to the generally recognized equation (\ref{fick1}). Surely, for the
case when $\theta$ is a linear function of $P$ and there is no
dissipation, equations (\ref{gen}), (\ref{GG}) are reduced to
(\ref{fick1}).

Use of modelling function for coverage profiles is very convenient for analysis of diffusion processes.
These functions are sufficiently exact and can be used for direct calculation of diffusion coefficients starting with
the Fick equation.

In the present paper we propose a few modelling functions for coverage profiles of Li adsorbed on Dy-Mo(112). They
described experimental data with a high precision, the deviation from experimental data is less than experimental
error. Thus the modelling functions present useful tools for both qualitative and quantitative studies of the
diffusion systems.

Let us note that calculating the modelling function for small
concentrations with higher accuracy (i.e., using all experimental
given points) it is possible to find a spike for the diffusion
coefficient for small concentrations.  The related plot is given by
Fig. 9.
\begin{figure}[th]
\centerline{\includegraphics[width=80mm]{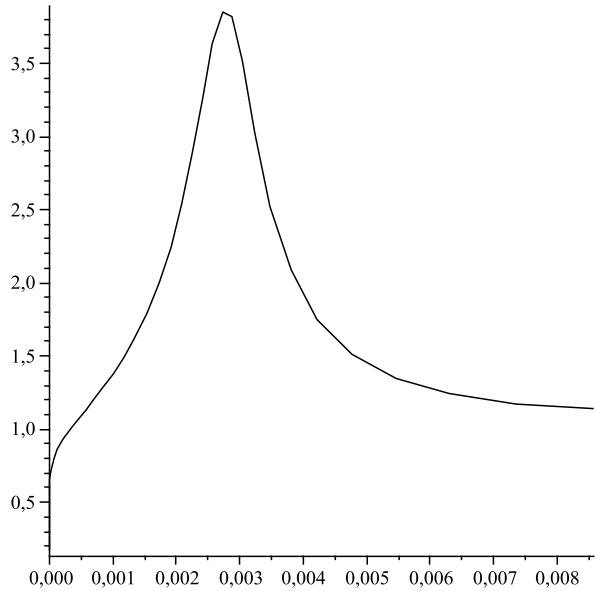}}
\caption{Diffusion coefficient $D(\theta)\ (\texttt{in units }
\texttt{mm}^2/\texttt{s}\cdot10^{-6})$ of Li on Dy-Mo(112)
calculated precisely for small concentrations. T=600 K.
 } \label{DG}
\end{figure}

Notice that the spike in Fig. 9 is indicated for a very small
concentrations compatible with the experimental error and so it
cannot be treated as well experimentally justified. On the other
hand this spike can be observed for all series of experimental data
presented by Fig. 2 and also in some other diffusion systems, e.g.,
in Dy absorbed on Mo(112). Indeed the coverage  profile given by
Fig. 8 is rather steady at $\xi=5-7$ when $\theta\sim 0.02$ and so
the related diffusion coefficient will have a maximum thanks to
small value of the derivative $\partial \theta/\partial \xi$.

Concerning interpretation the spike indicated by Figs. 9 and
min.-max. presented by Fig. 6 we can mention that the coverage
dependence of diffusivity is due to lateral interactions of
diffusing atoms and partially also to the specific features of the
substrate atomic structure (both intrinsic and caused by defects).
The combined action of lateral interactions and surface potential
corrugation determines the sequence of phase transitions that occur
in the diffusion zone. The phases can differ from one another not
only by the diffusion parameters, but also by the diffusion
mechanisms (see e.g. a review [8] and references therein). The
maxima of diffusivity at low coverage were observed for a number of
systems. We attribute this effect to a high mobility of single
adatoms. The fast decrease of D with growing coverage may be caused
by formation of clusters, whose diffusion mechanisms can be very
diversified, but their mobility is generally lower than that of
single adatoms. The diffusivity is also rather low in the regions of
first-order phase transitions in adlayers. As the coverage grows
further and the adlayer becomes increasingly dense approaching a
close-packed monolayer, the diffusion takes on a pronounced
collective character. In particular, in the region of
commensurate-incommensurate phase transition the diffusion seems to
proceed by a relay-race walks of misfit dislocations (topological
solitons), which provides a high diffusion rate (a local D maximum).
This effect was revealed for many adlayers. For more details refer
to papers \cite{naum}, \cite{naum1}.

Summarizing, we propose a constructive and convenient algorithm
(ERFEX) for generating of modelling functions which is valid for
arbitrary sufficiently smooth curves not necessarily related to a
diffusion process. In particular, using this algorithm and starting
with experimental data, it is possible to determine the diffusion
coefficient with a higher accuracy than the BM approach and spline
approximation. The algorithm can be treated as a specific
generalization of the wavelet approach which can be applied to study
of diffusions.

\begin{flushleft}
{\bf Acknowledgments}
\end{flushleft}
This work was supported by National Academy of Sciences of Ukraine under Project VC-138.

\begin{flushleft}
{\bf Appendix}
\end{flushleft}

Here we present the table including experimental data obtained for coverage profiles of Li adsorbed by Dy-Mo(112). They are used in the main text to estimate exactness of the modelling functions.

\newpage

\begin{center}
{Coverage profiles of Li adsorbed on Dy-Mo(112).}
\end{center}
\begin{tabular}{|c|c|c|c|c|c|}\hline
$x $ (mm)&$\theta_1,t=0$&$\theta_2, t=1200$ s&$\theta_3, t=2100$ s&$\theta_4, t=3600$ s&$\theta_5, t=5400$ s\\
\hline
0   &0.327&0.327&0.327&0.327&0.327\\
0.20&0.323&0.323&0.324&0.322&0.322\\
0.40 &0.322&0.323&0.319&0.318&0.316\\
0.48&     &     &     &     &0.310\\
0.56&     &     &     &     &0.300\\
0.60 &0.317&0.316&0.309&0.300&0.291\\
0.64&     &0.310&0.305&0.293&0.285\\
0.68&0.317&0.317&0.295&0.283&0.274\\
0.72&0.316&0.297&0.288&0.272&0.262\\
0.76&0.318&0.283&0.270&0.258&0.248\\
0.80 &0.312&0.266&0.253&0.239&0.234\\
0.84&0.299&0.232&0.221&0.217&0.215\\
0.88&0.201&0.193&0.194&0.194&0.195\\
0.90&0.067&0.165&     &0.177&0.181\\
0.92&0.036&0.137&0.155&0.159&0.169\\
0.94&0.025&0.103&0.131&0.144&0.154\\
0.96&0.013&0.066&0.104&0.122&0.140\\
0.98&0.010&0.040&0.074&0.105&0.124\\
1.00&0.008&0.018&0.051&0.081&0.106\\
1.02&&0.012&0.030&0.067&0.091\\
1.04&0.003   &0.008&0.018&0.049&0.077\\
1.08&0.001   &   0.003  &0.005&0.021&0.047\\
1.12 &0&0.002&0.003&0.009&0.027\\
1.16&0&0&        &      &0.011\\
1.2&0    &0&0       &0.001&0.004\\
1.28&&&&&0.001\\
\hline
\end{tabular}

 \end{document}